\documentclass[%
 reprint,
superscriptaddress,
aps,
prb,
 amsmath,amssymb,
]{revtex4-2}

\newcommand{\angstrom}{\textup{\AA}}
\setcitestyle{super}

\usepackage{graphicx, array,multirow }
\usepackage{dcolumn}
\usepackage{bm}
\usepackage{booktabs,xcolor,colortbl}
\usepackage{xparse}
\usepackage[toc,page]{appendix}
\usepackage{caption}
\usepackage{subcaption}
\usepackage{float}
\usepackage{afterpage}

\usepackage{tabularx,booktabs}

\usepackage{palatino}
\usepackage{hyperref}
\usepackage{todonotes}
\usepackage{color}

\begin{document}

\title{Database mining and first-principles assessment of organic proton-transfer ferroelectrics}

\author{Seyedmojtaba Seyedraoufi}
\author{Elin Dypvik Sødahl}
\affiliation{Department of Mechanical Engineering and Technology Management, Norwegian University of
Life Sciences, 1432 Ås, Norway.}

\author{Carl Henrik Görbitz}
\affiliation{Department of Chemistry, University of Oslo, 0371 Oslo, Norway.}

\author{Kristian Berland}
\affiliation{Department of Mechanical Engineering and Technology Management, Norwegian University of
Life Sciences, 1432 Ås, Norway.}
\email[E-mail: ]{kristian.berland@nmbu.no}

\date{\today}

\begin{abstract}
In organic proton-transfer ferroelectrics (OPTFe), 
molecules are linked together in a hydrogen-bonded network and proton transfer (PT) between molecules is the dominant mechanism of ferroelectric switching. 
Their fast switching frequencies 
make them attractive alternatives to 
conventional ceramic ferroelectrics, which contain rare and/or toxic elements,
and require high processing temperatures.
In this study, we mined the Cambridge Structural Database for potential \mbox{OPTFes}, 
uncovering all previously reported compounds, both tautomers and co-crystals, in addition to seven new candidate tautomers.  
The mining was based on identifying polar crystal structures with pseudo center-of-symmetry and viable PT paths. The spontaneous polarization and PT barriers were assessed using density functional theory. 
\end{abstract}

\maketitle


\section{\label{sec:level1} Introduction}
Organic ferroelectrics have several potential advantages over traditional ceramic ferroelectrics, including biocompatibility, printability, low weight, and flexibility.\cite{alaa, OFE, felx, biocompat} These properties make organic ferroelectrics a promising class of materials for application in wearable electronics.\cite{wearblee} Among these materials, organic ferroelectric polymers, such as polyvinylidene fluoride (PVDF), find use in various industries, including energy harvesting.\cite{pvdf1}
However, a key limitation of polymers is their high coercive fields\cite{alaa} which limits their applicability in low-voltage electronics.\cite{pvdf} 
Organic proton-transfer ferroelectrics (\mbox{OPTFe}), on the other hand, are characterized by their 
low coercive fields\cite{OFE} (on average $\sim 10~\mathrm{kV/cm} $) and fast ferroelectric switching.\cite{song, crca} 

Generally, one distinguishes between order-disorder ferroelectrics, in which ferroelectric switching involves the reorientation of static dipoles, and displacive ferroelectrics, in which ferroelectric switching comes from the repositioning of atomic species.\cite{book} 
In \mbox{OPTFes}, the polarization reversal is largely attributed to proton 
transfer (PT) from one
molecule to the next. 
In some cases, the switching mechanism is a combination of PT and displacement or reorientation of molecules, such as for the ionic molecular crystals of diazabicyclo[2.2.2]octane cations 
and tetrahedral inorganic anions.\cite{book, dabco, song} 
So far, the largest reported spontaneous polarization among the OPTFes stands at $\sim30~\mathrm{\mu C/cm^2}$ for croconic acid.
In comparison, a value of $\sim7~\mathrm{\mu C/cm^2}$ has been found for PVDF.\cite{crca, alaa}
Croconic acid, however, is corrosive, which can be problematic for device applications. Susceptibility to oxidation in the air is also another issue for some tautomeric ferroelectrics with $\rm OH \cdots O$ hydrogen bonds.\cite{rezbop}
Thus, it is important to identify broader types of \mbox{OPTFes}, for instance, tautomeric ferroelectrics with an imidazole ring are not corrosive and less prone to oxidation.\cite{rezbop}
Acid-base co-crystals are another type of \mbox{OPTFes} which typically have lower coercive fields, and spontaneous polarizations compared to the tautomeric ones. \cite{book}

In this paper, we present a systematic mining and screening study of all entries in the Cambridge Structural Database\cite{csd} (CSD) to identify new potential OPTFes.  
Spontaneous polarizations and PT barriers were computed 
using density functional theory (DFT)
for all seven compounds found in the mining, in addition to the compounds previously reported as OPTFes. 
DFT studies of OPTFes are scarce. Earlier works have been conducted by Lee et al. on the proton transfer barriers in the phenazine-chloranilic acid co-crystal \cite{chloranilic}, as well as by Ishibashi et al. on a set of tautomeric OPTFes. \cite{shoji}

This study is, in addition to identifying new candidate compounds, the first extensive study of PT barriers in OPTFes. We found that 
single barriers correlate well with the experimental coercive fields of earlier reported acid-base co-crystal systems.
For tautomers, on the other hand, 
collective barriers correlate quite well with the coercive fields, 
and can thus be used to provide a prediction of the coercive fields for the newly identified PT tautomers.

\begin{figure*}[t!]
\includegraphics[width=\textwidth]{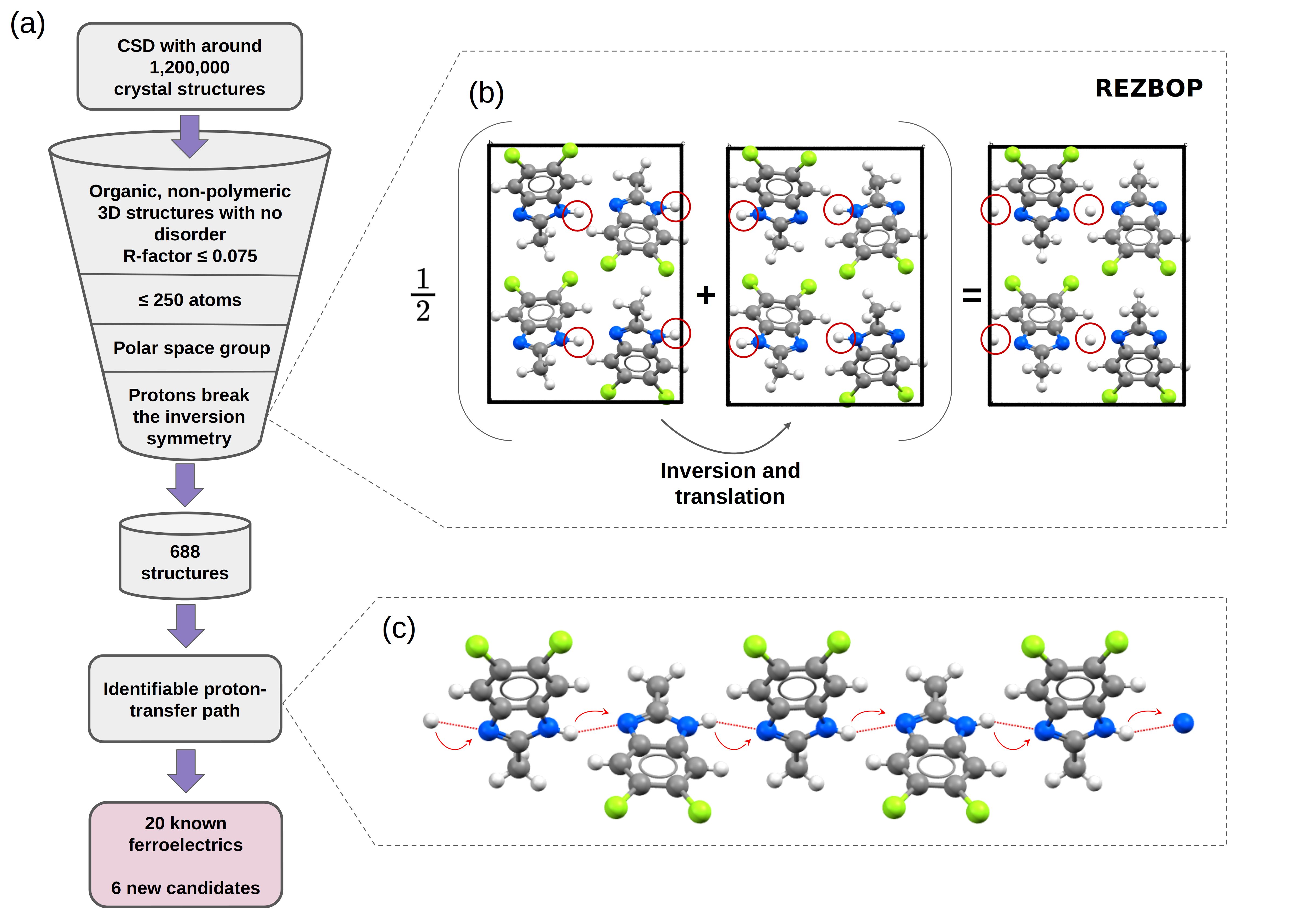}

\caption{\label{fig:workflow} 
(a) The workflow of the CSD mining. (b) Schematic illustration of inversion symmetry method on DC-MBI (CSD refcode: REZBOP), which is an OPTFe. The transferrable protons are marked with red circles. (c) The PT (hydrogen bond) paths are marked with red arrows. }
\end{figure*}

\begin{figure*}[t!]
\includegraphics[width=\textwidth]{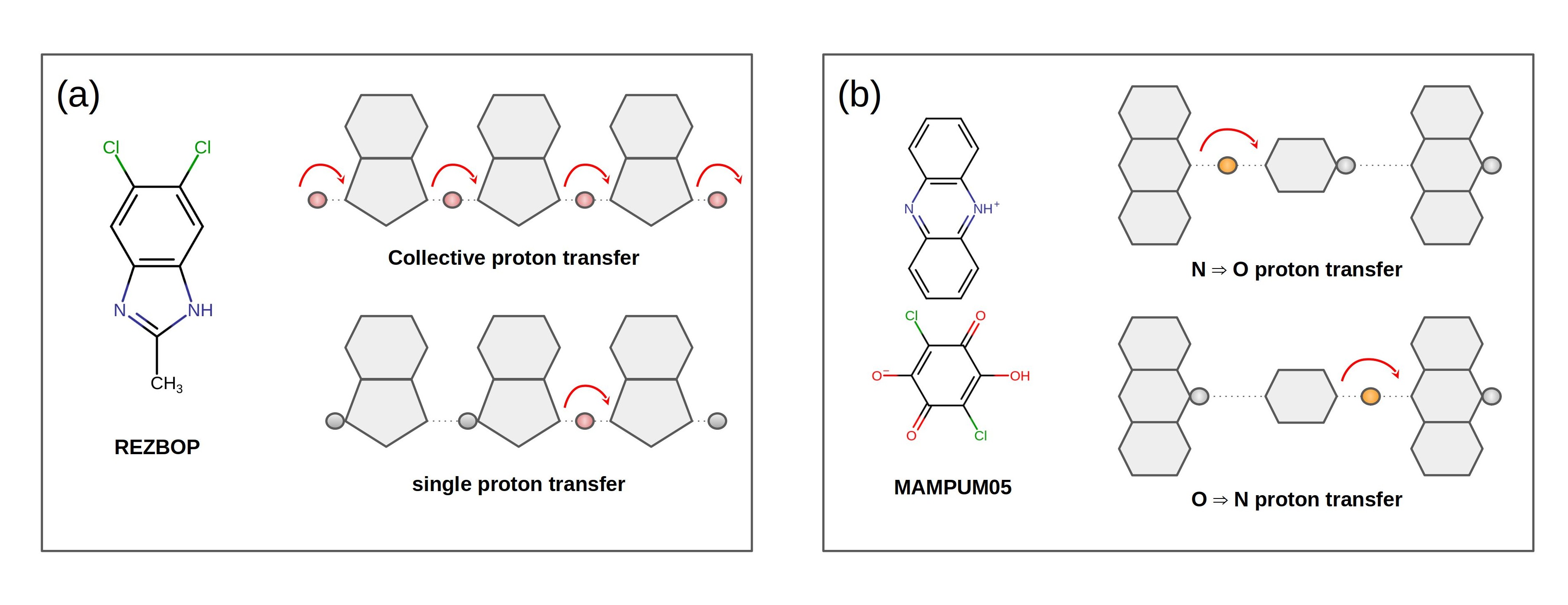}

\caption{\label{fig:schematic} 
(a) collective PT versus single PT is schematically shown for tautomers. The example system is DC-MBI (REZBOP). (b) Two possible single PT is illustrated for acid-base co-crystals. The example system is Phz-H2ca (MAMPUM03).}
\end{figure*}

\begin{figure*}[t!]
\includegraphics[width=\textwidth]{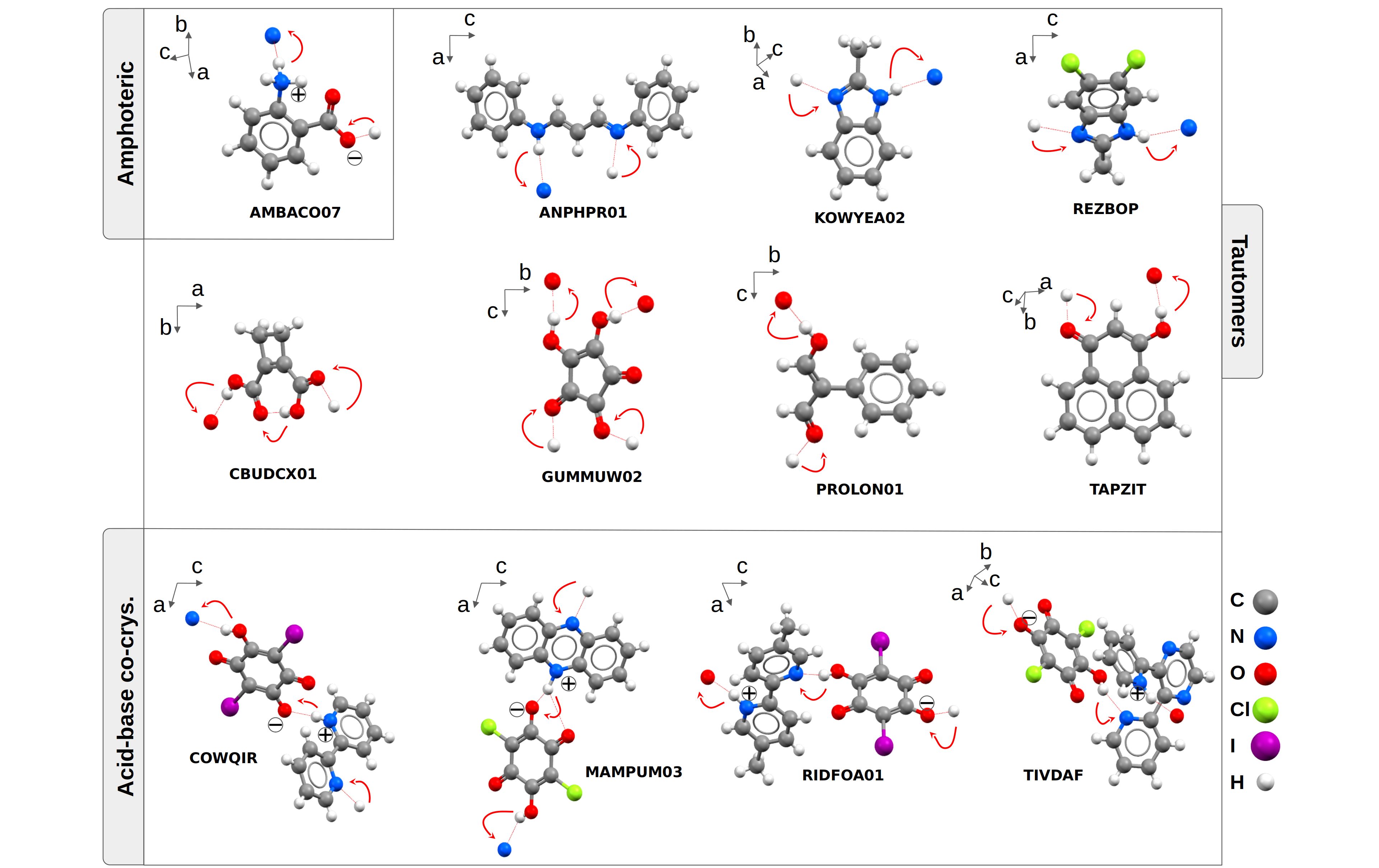}

\caption{\label{fig:dataset} 
Structures of the \mbox{OPTFes} benchmarking dataset. Arrows indicate the direction of site-to-site PT. COWRAK (not shown) has a similar structure as RIDFOA01, but Br replaced with I. Likewise, TIVFEL (not shown) is similar to TIVDAF, but Br replaces with Cl.}
\end{figure*}

\section{\label{sec:method}  Methods}
\subsection{\label{sec:csd_mine}CSD Searching approach}

Fig.~\ref{fig:workflow} panel (a) shows the workflow of the CSD mining.
In our screening study, we only included structures with a polar space group, which is a requirement for ferroelectricity\cite{sym_FE},
and limited our investigation to organic non-polymeric structures with fewer than 250 atoms. Filters were applied for determined 3D coordinates, \mbox{R-factor} $\leq$~0.075, and no structural disorder. This reduced the dataset from more than $\sim$~1,200,000 to $\sim$~80,000 crystal structures. The CSD Python API 3.0.11\cite{api} was used to implement these criteria. \par
Next, from this pool, we used the Python-based tool \textsc{Molcrys}\cite{molcris}, building on the Atomic Simulation Environment (\textsc{ASE})\cite{ASE} and
\textsc{networkx}\cite{networkx}
to exclude all compounds lacking an approximate pseudo-center of symmetry. Pseudosymmetry refers to the concept where a structure can be viewed as a higher symmetry structure that has undergone minor distortions. The idea to find ferroelectric materials through pseudosymmetry detection has been proposed\cite{pseduo} and explored in previous studies: Capillas et al.\cite{Capillas} suggested a procedure to detect pseudosymmetry by searching through the minimal supergroups of the material's space group. Abrahams\cite{Abrahams} exploited the same idea to detect new inorganic ferroelectrics. 
A key criterion for prospective OPTFe is that the structures without hydrogens
should have pseudo-center of symmetry. 
 
The inset {\bf (b)} of Fig.~\ref{fig:workflow} outlines 
the procedure to find the structures with pseudo-center of symmetry. 
The example crystal structure is a tautomeric OPTFe with a clear nitrogen-to-nitrogen PT path, as is shown in the panel {\bf (c)}. 
Inversion of the crystal structure, followed by a translation to overlay the molecules based on minimizing the mean deviation of the atomic separation between equivalent atoms in the images, results in a close-to-identical structure except for protons involved in the PT (highlighted with red circles).

Once structures with pseudo-center of symmetry and "isolated protons" in the middle of hydrogen bonds have been identified, we inspected the crystal structures to identify a clear sequence of PT pathways that allow a cascade of PTs, as opposed to for instance intra-molecular PT or inter-molecular PT within supramolecular complexes not connected by PT paths. 
Panel {\bf (c)} shows a case of connected paths.

\subsection{\label{sec:dft_method}Density functional theory calculations}
The DFT calculations were performed with the projector augmented wave (PAW) method implemented in the \textsc{VASP} software package\cite{vasp, vasp1, vasp2} using hard PAW pseudopotentials.  
The plane-wave cutoff was set to 1200~eV, based on our earlier work on PT-barriers of molecular dimers.\cite{seyed} A high cut-off and hard pseudopotentials were also necessary to converge the lattice parameters for the \mbox{OPTFes} systems.
The self-consistent loop was iterated until energy differences reached below $10^{-8}$~eV. The Brillouin zone was sampled with a $\Gamma$-centered Monkhorst-Pack grid with ($1/25) \, \angstrom^{-1}$ spacing. 
This converged the lattice constants of croconic acid, the smallest unit cell volume in our set, to within $0.01 \, \angstrom$. 
The spontaneous polarization was computed using the Berry-phase method.\cite{berry_1, berry_2, berry_3} 

The climbing image nudged elastic band method\cite{ciNEB} was used to identify the intermolecular PT transition states in the acid-base co-crystals. 
However, for tautomers, the molecular states with both 
protons transferred onto one molecule were in general unstable. Thus, the single PT barriers for these compounds were not computed. 
For collective PT, the transition state was approximated using the same interpolated image used to identify the pseudo-center of symmetry, as illustrated in Fig.~\ref{fig:workflow}{\bf (b)}, followed by a single-point DFT evaluation.
The difference between the single and collective PT are schematically illustrated in Fig.~\ref{fig:schematic}{\bf (a)} for the tautomer REZBOP.
During the single PT, one proton is moved along the $\rm NH \cdots N$ hydrogen bond to the adjacent molecule. However, in collective motion, all protons are moved simultaneously. 
The full collective barrier scales with the size of the crystal,
but it does give an indication of whether many protons are
involved in the ferroelectric switching, 
and the corresponding ferroelectric grain boundaries.

For the acid-base co-crystals, 
we have two types of molecules, and, as a result, the single PT can happen in two possible ways, 
i.e., from nitrogen to oxygen and vice versa, as depicted in Fig.~\ref{fig:schematic}{\bf (b)}. In making our correlation analysis, we assumed the lowest barrier was the rate-limiting and used this for correlation analysis. 

Piezoelectric properties were computed using density functional perturbation theory.\cite{gajdos_linear_2006, baroni_ab_1986} 

\section{\label{sec:results}  Results and discussion}
\subsection{\label{sec:results_1}Earlier reported OPTFes} 
\begin{figure}[t!]
\includegraphics[width=\columnwidth]{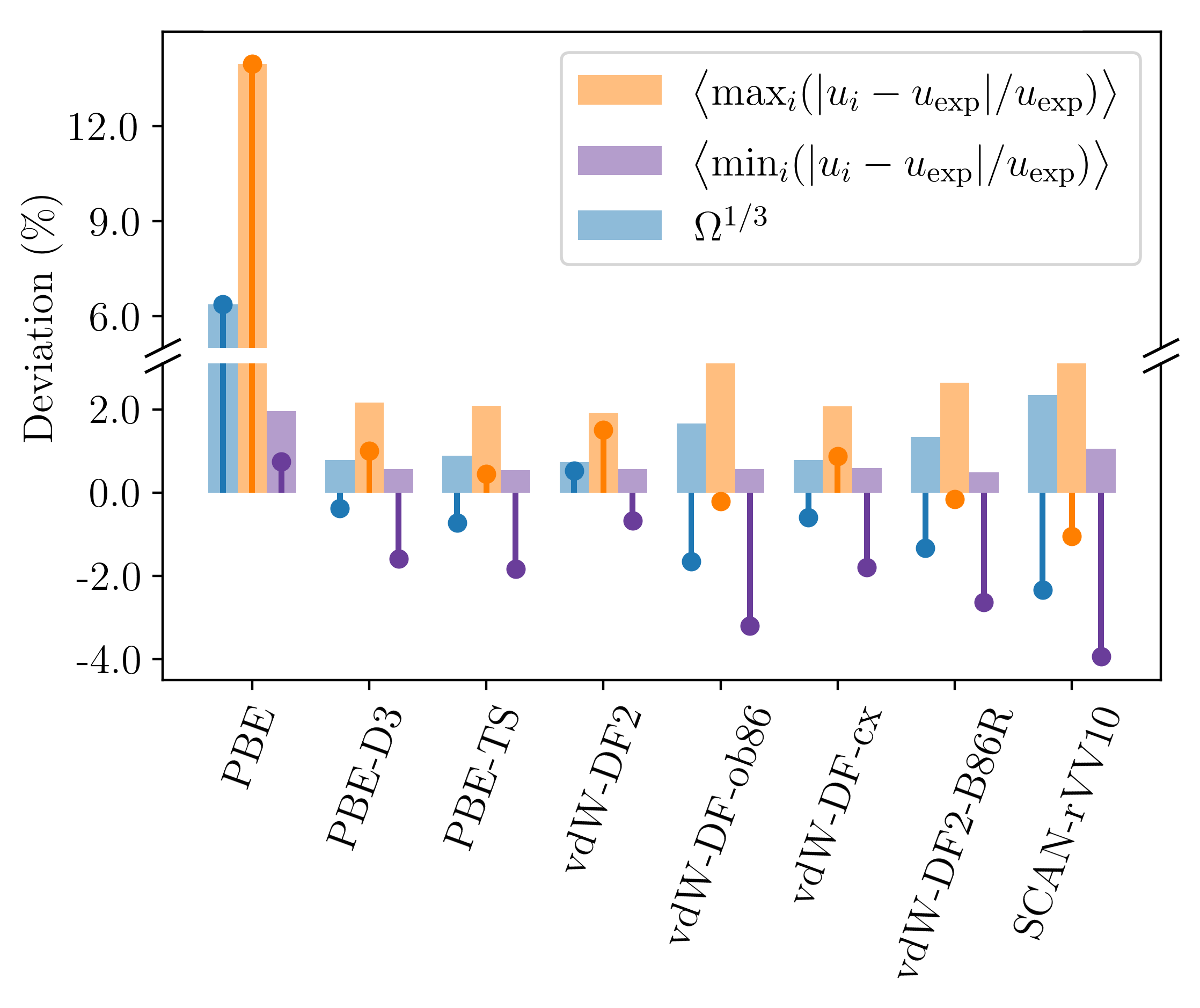}

\caption{\label{fig:functionals} 
Mean absolute relative deviation of the third root of the volumes $\Omega^{1/3}$, and the mean of the largest (and smallest) absolute relative deviation of
lattice constants ($u_i$) depicted in light-color bars. The dark-color bars are corresponding relative (not-absolute) deviations. }  
\end{figure}
\begin{figure}[t!]
\includegraphics[width=\columnwidth]{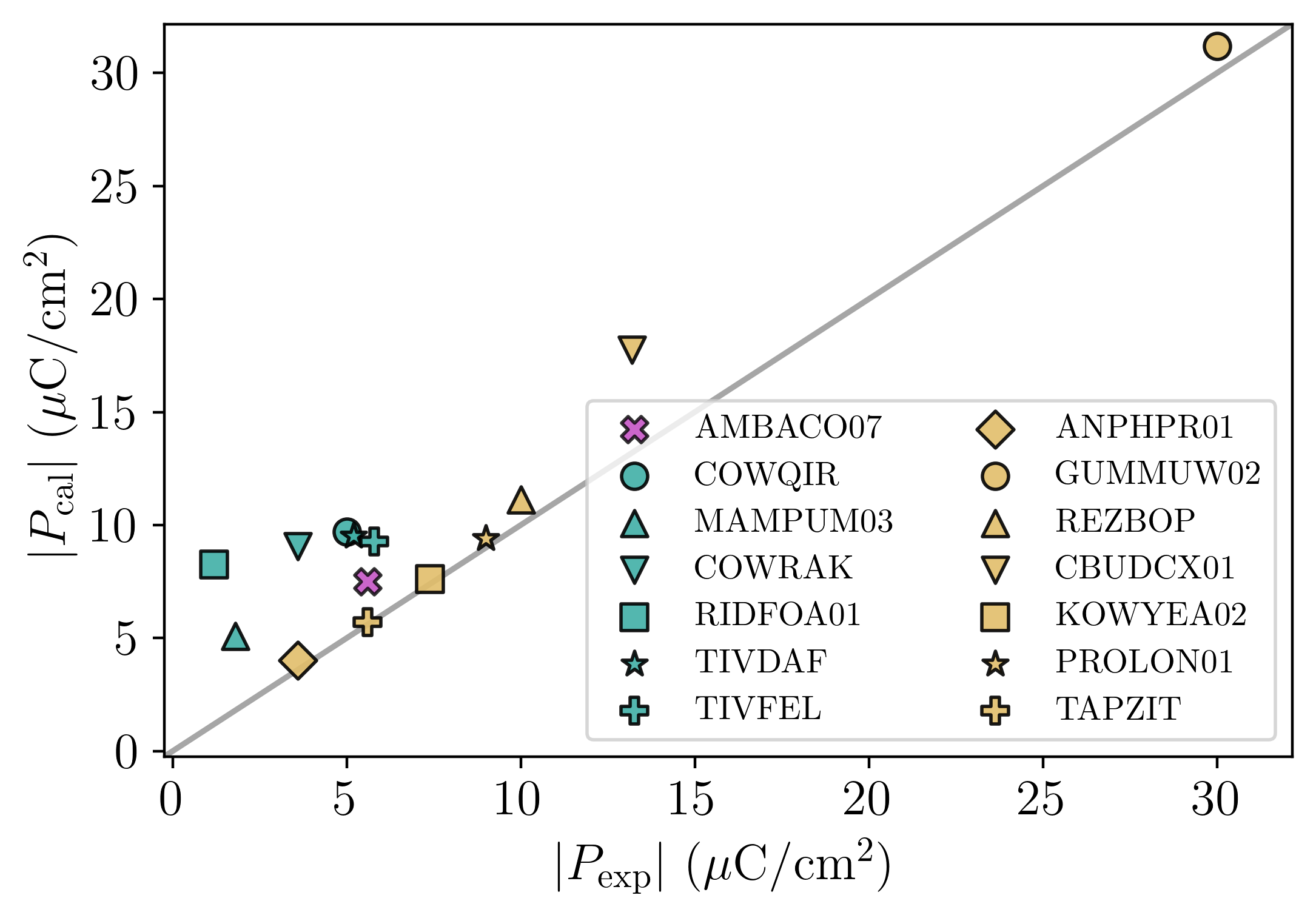}

\caption{\label{fig:pol_cal_exp} 
Computed versus experimental spontaneous polarization for the \mbox{OPTFes} dataset. }
\end{figure}

\begin{figure}[t!]
\includegraphics[width=\columnwidth]{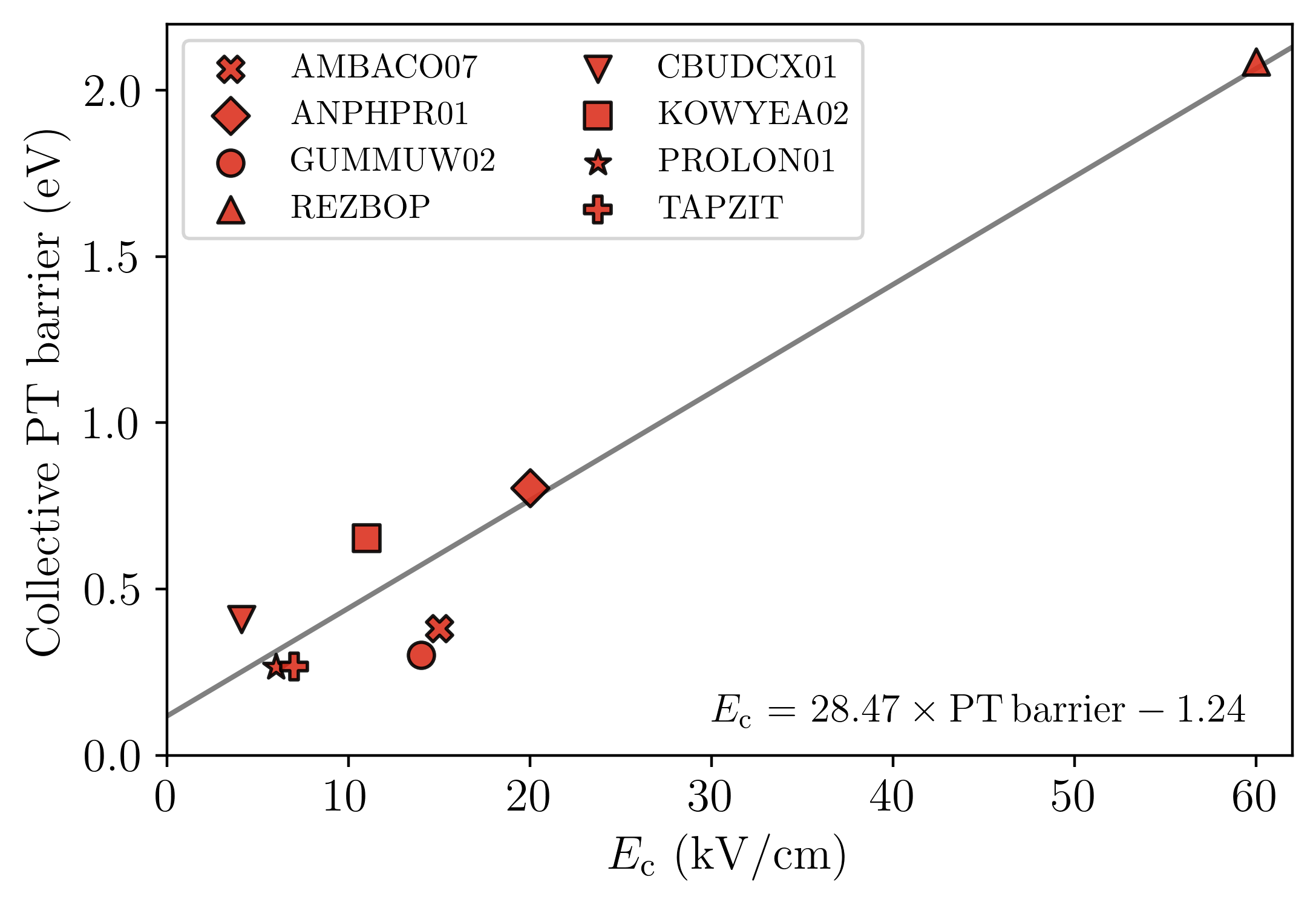}
\includegraphics[width=\columnwidth]{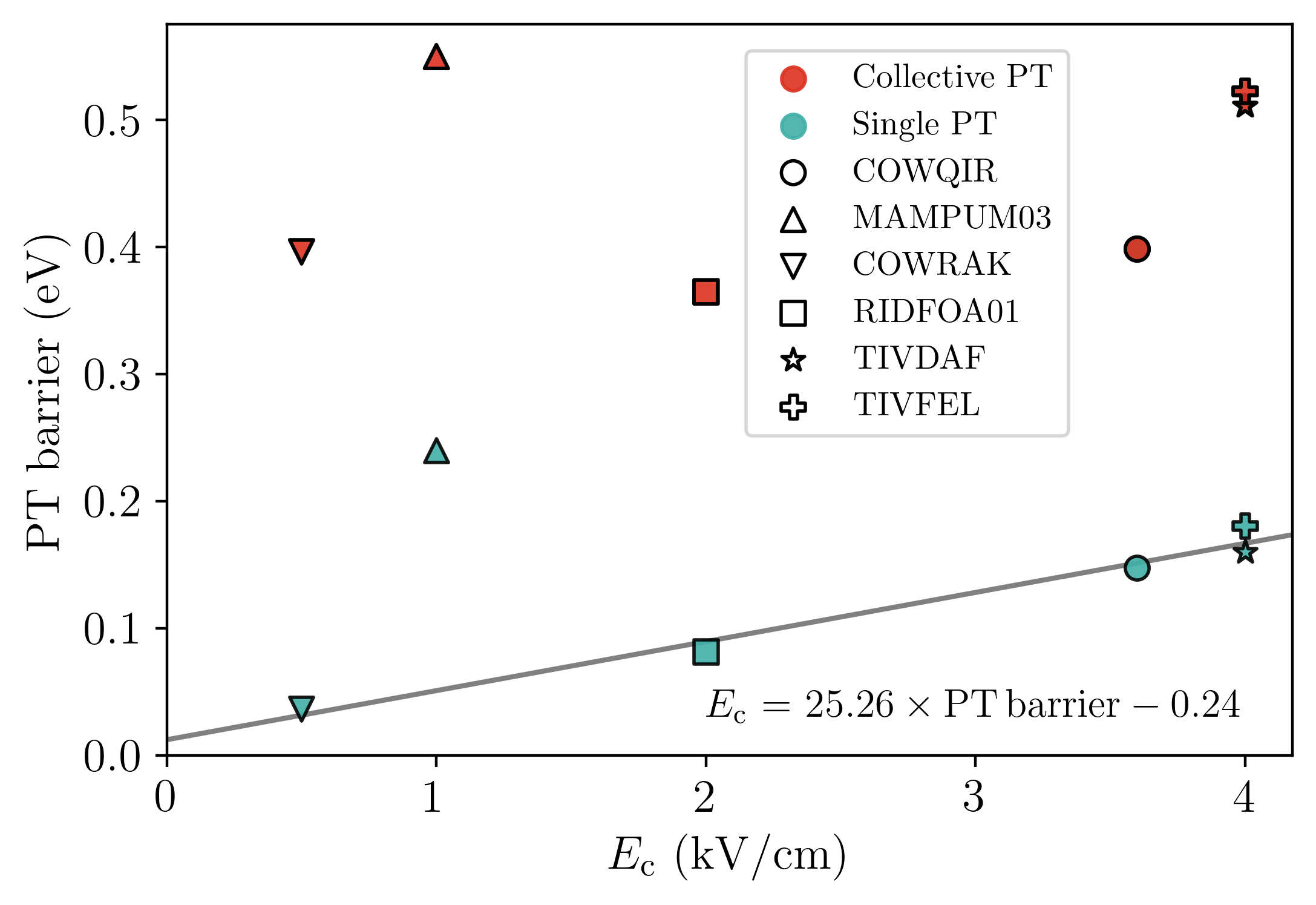}
\caption{\label{fig:Ec_predict} 
In the upper panel, the collective PT energy barrier per proton is plotted against the coercive field ($E_{\rm c}$) for the tautomeric \mbox{OPTFes}. AMBACO07 which is the amphoteric type is also shown in the panel.
In the lower panel, collective and single PT energy barrier per proton is plotted for acid-base co-crystal \mbox{OPTFes}. }
\end{figure}

\begin{figure*}
\includegraphics[width=\textwidth]{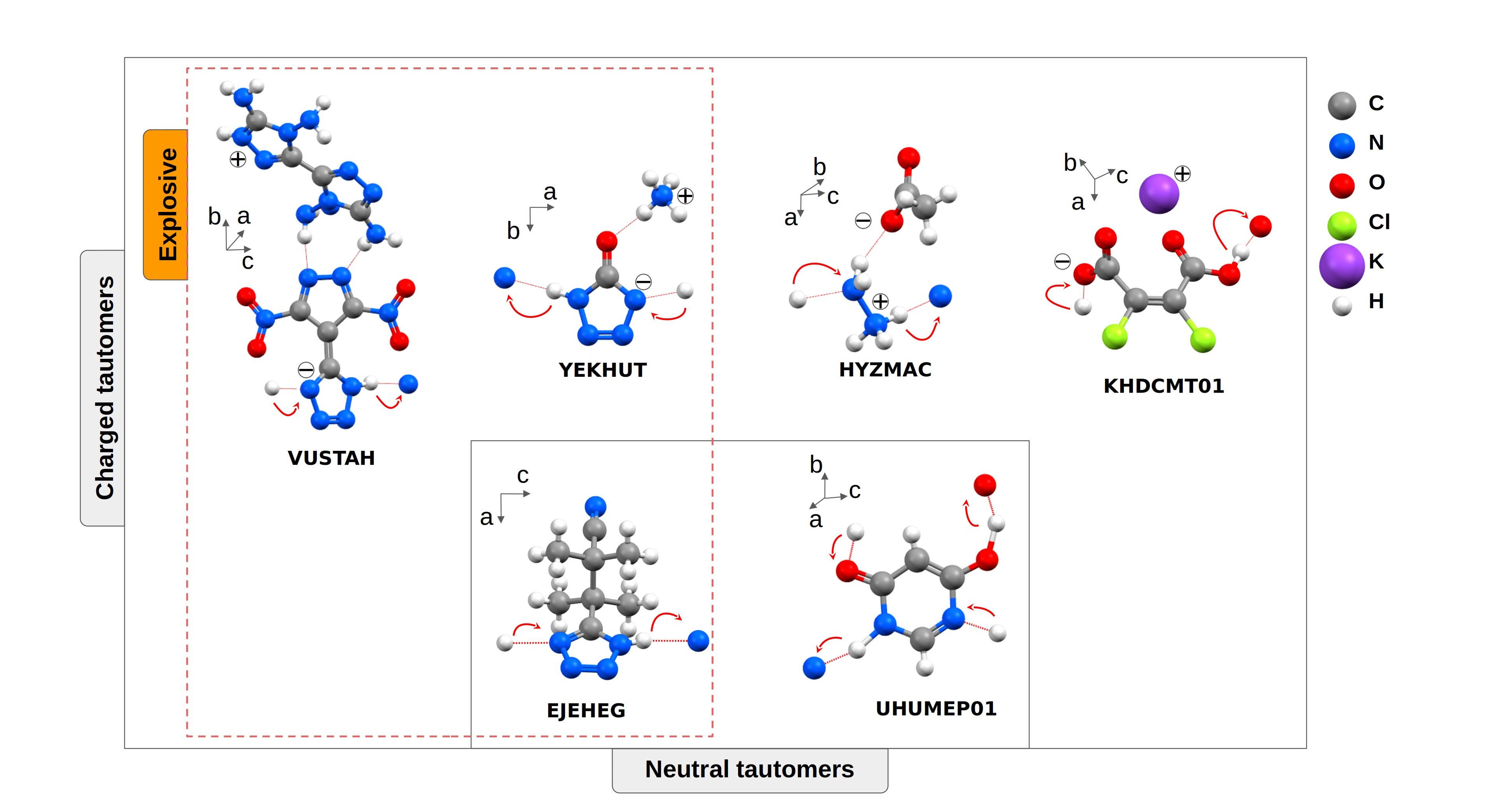}
\caption{\label{fig:systems} 
Structures of the identified \mbox{OPTFes} candidates. Arrows indicate the direction of site-to-site PT. Explosive materials are indicated by an orange box.}
\end{figure*}
There are more than 20 entries for known OPTFes in the CSD, including structures of acid-base and of single molecules that display proton tautomerism.\cite{book,alaa, antharnilic, cbdc,tivdaf,cowqir, crca, dabco,mampu, rezbop, ridfoa,song}
To compare experimental and computed values, 
we defined a benchmark set comprising seven tautomers, one amphoteric compound, and six acid-base co-crystals,
for which we computed lattice parameters with several exchange-correlation functionals, as well as spontaneous polarization and PT barriers. For the structures that were experimentally resolved at various temperatures, we selected the one with the lowest temperature. Lattice parameters can be very sensitive to the choice of exchange-correlation functional, in particular for vdW-bonded systems.\cite{vdw_3, facingchall} 
Fig.~\ref{fig:dataset} shows the molecular species and PT paths. 
In this paper, we refer to the structures by their CSD refcodes. 
The corresponding chemical abbreviations are listed in Tab.~\ref{tab:table_1} and Tab.~\ref{tab:table_2}.

In the amphoteric compound AMBACO07, the hydrogen-bonded one-dimensional network of zwitterionic and neutral anthranilic acid molecules enables ferroelectric switching.\cite{antharnilic}

Among the known tautomers, PT occurs 
between a couple of different moieties, as shown in Fig.~\ref{fig:dataset}.  
In the ANPHPR01, KOWYEA02, and REZBOP crystals, the transfer is between enamine–imine moieties.  
In CBUDCX01, the transfer is between carboxylic acid (HO-C=O) moieties, and PROLON01 and TAPZIT have HO-C=C-C=O moieties. 
Croconic acid (GUMMUW02) has a double PT moiety HO-C=C-OH-C=O-C=O 
forming a three-dimensional network of PT pathways.\cite{rezbop, crca, cbdc, alaa} 
The acid-base co-crystals are all composed of halogenated anilic acid derivatives combined with either bipyridine (COWQIR), phenazine (MAMPUM03), dimethyl-bipyridine (RIDFOA01) or pyridine-pyrazine (TIVDAF) derivatives.\cite{cowqir, mampu, ridfoa, tivdaf} 

\subsubsection{\label{sec:results_2} Benchmarking of lattice constants}

Fig.~\ref{fig:functionals} compares the accuracy of the lattice vectors of the benchmark set with the experimentally reported structures for different functionals.  
Among the \mbox{vdW-DF} density functionals, \mbox{vdW-DF2}\cite{vdw2} and \mbox{vdW-DF-cx}\cite{cx} have the best agreement with the experimental lattice constants, the latter is in line with the results of Ishibashi et al.\cite{shoji} 
In contrast, PBE\cite{pbe} overestimates and \mbox{SCAN-rVV10}\cite{scan_rvv10} significantly underestimates lattice constants. \mbox{PBE-D3}\cite{dft_d3} and \mbox{PBE-TS}\cite{pbe-ts}, which use atom-centered vdW-dispersion correction to account for dispersion forces, 
both provide accurate lattice parameters. In an earlier study, we found that the use of non-local correlation at the vdW-DF level greatly improved PT barriers for a set of molecular dimers and that \mbox{vdW-DF2}, in particular, provided accurate PT barriers
\cite{seyed}.
Thus, we adopted vdW-DF2 in this study.

\subsubsection{\label{sec:prop_1}Ferroelectric properties}

Fig.~\ref{fig:pol_cal_exp} plots the computed, against the reported experimental 
spontaneous polarization values, which are also 
listed in Tab.~\ref{tab:table_1} and Tab.~\ref{tab:table_2}. 
For tautomers, the computed and experimental values agree within 10~$\%$.
However, for co-crystals, the computed values overestimate the experimental values on average by a factor of three. 
Kagawa et al.\cite{domain_wall} reported that domains can be strongly pinned in OPTFe co-crystals, causing a discrepancy between computed and experimental values. In their case, with additional thermal annealing, the spontaneous polarization of 6,6'-dimethyl-2,2'-bipyridinium chloranilate, an OPTFe co-crystal, increased from 1.3 $\rm \mu$C/cm$^2$ to 7 $\rm \mu$C/cm$^2$.

\begin{table*}[t!]
\caption{\label{tab:table_1} \raggedright PT energy barriers, and ferroelectric properties of reported acid-base co-crystal \mbox{OPTFes}. $|P_{\rm exp}|$ and $|P_{\rm cal}|$ are experimental and computed spontaneous polarizations ($\rm \mu$C/cm$^2$) respectively. $E_{\rm c}$ (kV/cm) is the coercive field and $T_{\rm C}$ is the Curie temperature. $\Delta_{\rm }^{\rm col}$ is collective PT and $\Delta_{\rm }^{\rm sgl}$ is single PT energies. }
\begin{ruledtabular}
\begin{tabular}{lccccccccc}
CSD refcode & $T_{\rm C}$ (K) & $E_{\rm c}$ & $|P_{\rm exp}|$ &  $|P_{\rm cal}|$ & $\Delta_{\rm O \rightarrow N}^{\rm sgl}$~(eV) & $\Delta_{\rm N \rightarrow O}^{\rm sgl}$~(eV) & $\Delta_{\rm }^{\rm col}$~(eV) & Space group &  Chemical abbreviation\\
\hline
\midrule

COWQIR & 158 & 3.6 & 5.0 & 9.70 & 0.15 & 0.24 & 0.40 & $\rm P1$ & H22bpy-Hia \cite{cowqir} \\
MAMPUM03 & 253 & 1.0 & 1.8 & 5.09 & 0.46 & 0.24 &  0.55 & $\rm P2_1$ &  Phz-H2ca \cite{mampu} \\
COWRAK & 259 & 0.5 & 3.6 & 9.06 & 0.04 & 0.22 & 0.40 & $\rm P1$ &  H55dmbp-Hba \cite{cowqir} \\
RIDFOA01 & 268 & 2.0 & 1.2 & 8.26 & 0.08 & 0.26 & 0.32 & $\rm P1$  & H55dmbp-Hia \cite{ridfoa}\\
TIVDAF & 402 & 4.0 & 5.2 & 9.50 & 0.16 & 0.33 & 0.52 &  $\rm Cc$   &  Hdppz-Hca \cite{tivdaf} \\
TIVFEL & 420 &  4.0 & 5.8 & 9.26 & 0.18 & 0.34 &  0.48 & $\rm Cc$ & Hdppz-Hba \cite{tivdaf}\\

\end{tabular}
\end{ruledtabular}
\end{table*}

\begin{table*}[t!]
\caption{\label{tab:table_2} \raggedright PT energy barriers, and ferroelectric properties of reported and potential OPTFes identified here. $|P_{\rm exp}|$ and $|P_{\rm cal}|$ are experimental and computed spontaneous polarizations ($\rm \mu$C/cm$^2$) respectively. $E_{\rm c}$ (kV/cm) is the coercive field, $T_{\rm C}$ is the Curie temperature, and $\Delta_{\rm }^{\rm col}$ is collective PT barriers. ($\ast$): Estimated $E_{\rm c}$ }.
\begin{ruledtabular}
\begin{tabular}{lcccccccc}
&CSD refcode & $T_{\rm C}$ (K) & $E_{\rm c}$ & $|P_{\rm exp}|$ &  $|P_{\rm cal}|$ & $\Delta_{\rm }^{\rm col}$~(eV) & Space group &  Chemical abbreviation\\
\hline
\midrule
&AMBACO07 & 357 & 15.0 & 5.6 & 7.50  & 0.38 & $\rm Pna2_1$ &  Anthranilic acid \cite{antharnilic} \\
\midrule
{\multirow{7}{*}{\rotatebox[origin=c]{90}{Tautomers}}} & ANPHPR01 & 348 & 20.0 & 3.6 & 4.03 & 0.80 & $\rm Iba2$ &  ALAA \cite{alaa} \\
&GUMMUW02 & 380 &  14.0 & 30.0 & 31.20  & 0.30 & $\rm Pca2_1$ &  Croconic acid\cite{crca} \\
&REZBOP & 399 & 60.0 & 10.0 & 11.14 & 2.08 & $\rm Pca2_1$ & DC-MBI  \cite{rezbop} \\
&CBUDCX01 & 400 & 4.1 & 13.2 & 17.75  & 0.41 & $\rm Cc$ & CBDC \cite{cbdc} \\
&KOWYEA02 & 449 & 11.0 & 7.4 & 7.61 &  0.65 & $\rm Pc$ &  MBI \cite{crca} \\
&PROLON01 & 363 & 6.0 & 9.0 & 9.42  & 0.26 & $\rm Pna2_1$ &  PhMDA \cite{cbdc} \\
&TAPZIT & 510 & 7.0 & 5.6 & 5.71  & 0.27 & $\rm Pc$ & 3HPLN \cite{cbdc} \\
\midrule
 \multicolumn{9}{c}{Identified PT ferroelectric candidates} \\
\midrule

{\multirow{6}{*}{\rotatebox[origin=c]{90}{Tautomers}}} 
&HYZMAC & -- &  16.47~$^{\ast}$ & -- & 14.90  & 0.65 & $\rm Cc$ & HAA\cite{hyzmac}\\
&EJEHEG & -- & 10.93~$^{\ast}$ & -- & 7.38  & 0.46 & $\rm Pca2_1$ &  3MTB \cite{EJEHEG} \\
&KHDCMT01 & -- & 1.17~$^{\ast}$ &  -- & 6.71  & 0.12 & $\rm P1$ &  PHDCM\cite{KHDCMT01} \\
&UHUMEP01 & -- & 8.62~$^{\ast}$ & -- & 4.41   &  0.38 & $\rm Cc$ & 6HP\cite{UHUMEP01_ref} \\
&VUSTAH & -- &  10.04~$^{\ast}$ & -- & 0.33  & 0.43 & $\rm Pna2_1$ & DADT-DNPT\cite{vustah} \\
&YEKHUT & -- & 18.00~$^{\ast}$ & -- & 8.39   & 0.70 & $\rm Fdd2$ &  A5OT\cite{yekjef} \\

\end{tabular}
\end{ruledtabular}
\end{table*}

As ferroelectric switching occurs through PT, 
the barriers should correlate with the corresponding coercive fields. 
Such a correlation for the tautomers is shown in the upper panel of Fig.~\ref{fig:Ec_predict}, which plots the experimental coercive fields against collective PT barriers. 
This comparison indicates that there is an approximately linear relationship between the calculated collective PT barriers and the measured coercive fields,
as presented by the regression line, which is obtained using the least squares method.

For the acid-base co-crystals, a linear relationship between the single PT barriers and coercive fields was found, as shown by a regression line in the lower panel of Fig.~\ref{fig:Ec_predict}. The single PT barriers are on average 70~$\%$ lower than the collective ones. \par
The lower single barriers and the linear relationship with the coercive field strongly suggest that ferroelectric switching has a single PT nature in the acid-base co-crystals, in contrast, to a degree of collectively in the polarization switching of tautomers. 
In acid-base co-crystals, nitrogen-to-oxygen single PT barriers ($\Delta_{\rm N \rightarrow O}^{\rm sgl}$) are in the most cases much larger than oxygen-to-nitrogen barriers ($\Delta_{\rm O \rightarrow N}^{\rm sgl}$). An exception is MAMPUM03 for which $\Delta_{\rm O \rightarrow N}^{\rm sgl}$ is almost twice that of $\Delta_{\rm N \rightarrow O}^{\rm sgl}$.

\subsection{\label{sec:new_cand}Identified OPTFe candidates}

The OPTFe candidates found in the screening are shown in Fig.~\ref{fig:systems}. 
The majority of the structures exhibit proton tautomerism. The exception is hydrazinium acetate HYZMAC, which lacks the rearrangement of $\pi$ orbitals, in which $\rm NH_3^{+} \cdots NH_2$ hydrogen bond forms potential switchable PT paths, with acetate anions as shielding spectators.  
Among the rest, 
EJEHEG is one of the three tetrazole-based candidates and has only a one-dimensional hydrogen bond network with $\rm NH \cdots N$ interactions running along the c-axis. As indicated by the red arrows, PT can potentially switch the polarization direction. 
KHDCMT01 is a hydrogen dichloromaleate potassium salt with zigzag $\rm COO^{-} \cdots COOH$ PT chains running along the ac-diagonals.  
UHUMEP01 is an analog of the RNA nucleobase uracil. The molecules can participate in a lactam-lactim tautomerism with two-interconnected and intersecting PT paths in the bc-plane with $\rm NH \cdots N$ and $\rm OH \cdots O$ hydrogen bonds, respectively.\cite{UHUMEP01_ref} 
YEKHUT is a co-crystal of 5-oxotetrazolate, and potentially explosive due to its high nitrogen content.\cite{yekjef} It forms crystals with orthorhombic symmetry, 16 ion pairs in the unit cell, and 3D hydrogen bonding. PT could occur in two sets of $\rm NH \cdots N$ chains in the ac plane.
Finally, VUSTAH is the last candidate with tetrazole backbone.\cite{vustah} Again, there is extensive hydrogen bonding, but PT may occur along simple $\rm NH \cdots N$ chains parallel to the 4.88 Å unit cell axis. 

In our screening process, we found another co-crystal of 5-oxotetrazolate (CSD refcode: YEKJEF\cite{yekjef}). However, we observed that the inversion symmetry was disrupted not only by the locations of the protons but also by the displacement of hydroxylammonium ions. As a result, YEKJEF was excluded from the list of identified candidates, but it could potentially serve as an interesting ferroelectric candidate with a mixed PT and molecular reorientation mechanism.  

\subsubsection{\label{sec:prop_2}Ferroelectric properties}
In the lower panel of Tab.~\ref{tab:table_2}, we have listed the computed spontaneous polarizations and PT barriers for the OPTFe candidates identified here. 
The values show a large variation, ranging from 0.3~$\rm \mu$C/cm$^2$ of VUSTAH to 14.9~$\rm \mu$C/cm$^2$ of HYZMAC. 
The large computed spontaneous polarization of HYZMAC can be related to the small area $\sim$~77~$\angstrom^2$ of the intersecting plane perpendicular to the polarization vector, compared with $\sim$~329~$\angstrom^2$ of VUSTAH. \par
\begin{figure}[]
\includegraphics[width=\columnwidth]{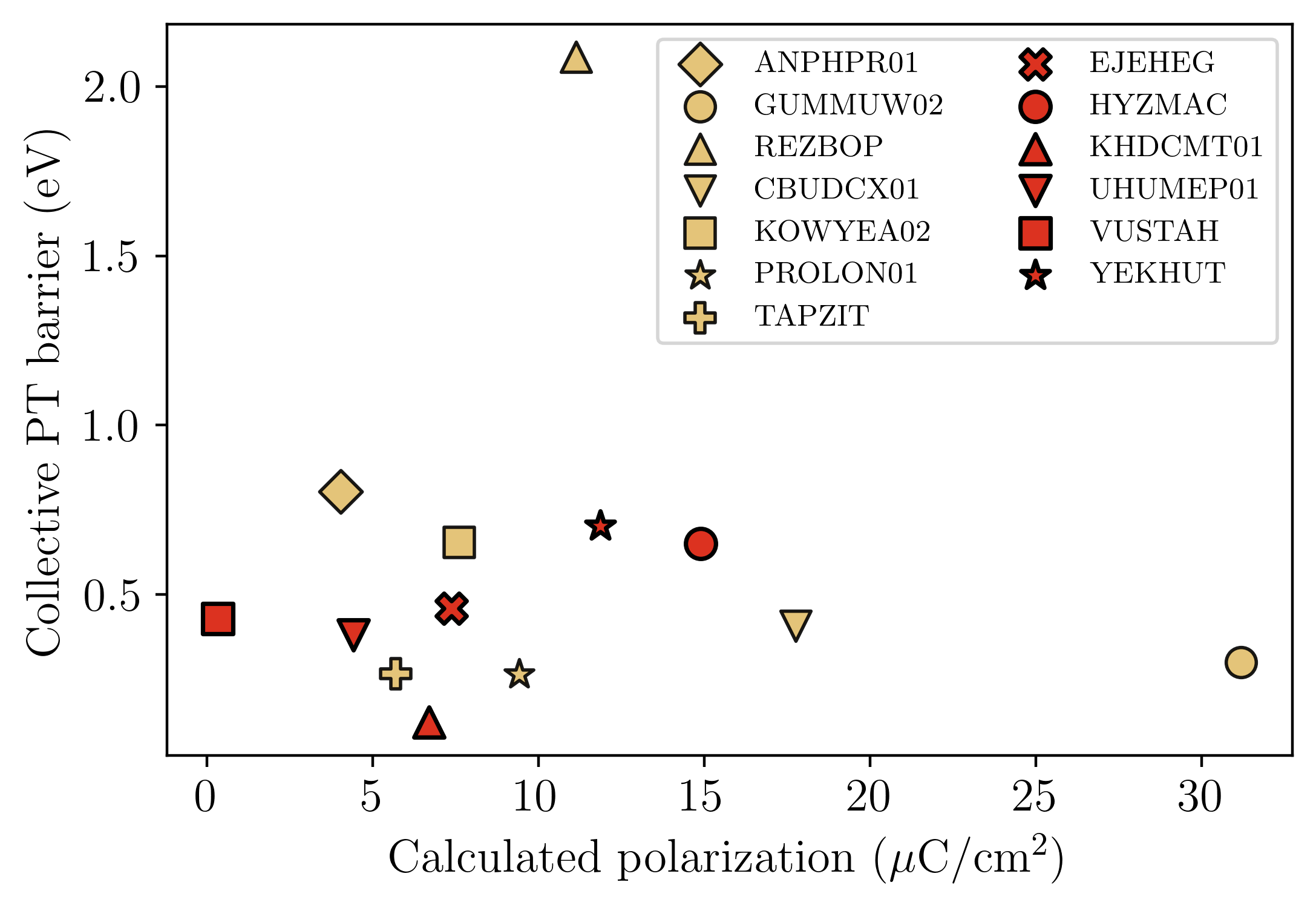}
\caption{\label{fig:pol_ec} 
Collective PT barriers plotted against the computed spontaneous polarizations for tautomeric systems and identified candidates.}
\end{figure}

Collective PT barriers were also computed for the identified candidates. Fig.~\ref{fig:pol_ec} plots the calculated polarization against the collective PT barriers for identified candidates and, for comparison, the earlier known candidates. All compounds except REZBOP have PT barrier, $\Delta^{\rm col}$, less than 1.0 eV, and except for croconic acid, all spontaneous polarizations are less than 20~$\rm \mu$C/cm$^2$ 
Because all of our identified ferroelectric candidates are tautomers, we used the linear correlation line for the existing tautomers to predict the coercive fields for the new ones, which fall in the range of 1 to 20 kV/cm as listed in Tab.~\ref{tab:table_2}.  

\begin{table*}
\caption{\label{tab:table_piezo} \raggedright Piezoelectric properties of candidate PT materials. }
\begin{ruledtabular}
\begin{tabular}{ccccccc}
 & EJEHEG & HYZMAC & KHDCMT01 & UHUMEP01&  VUSTAH & YEKHUT\\
\hline
$\mathrm{d_{31}}$ [pC/N] & -0.05 &  0.70 & -0.41 & 0.79 & -1.47 & -1.66 \\
$\mathrm{d_{33}}$ [pC/N] &  1.14 &  0.66 &  0.35 & 0.31 &  2.62 &  3.13 \\
$\mathrm{d_{25}}$ [pC/N] &  3.69 & -1.25 &  0.49 & 0.67 & -5.32 &  -60.3\\
$\mathrm{d_{16}}$ [pC/N] &  0.82 & -2.52 & -0.59 & 0.65 & -1.64 &  -4.39\\
$\epsilon_{r0,11} (\epsilon_{r\infty,11})$ & 2.77 (2.48) & 4.57 (2.50) & 4.54 (2.46) & 2.56 (2.01) & 4.77 (2.84) & 5.38 (2.53)  \\
$\epsilon_{r0,22} (\epsilon_{r\infty,22})$ & 2.70 (2.44) & 4.07 (2.47) & 4.29 (2.89) & 4.19 (2.95) & 4.56 (3.15) & 5.48 (2.81)  \\
$\epsilon_{r0,33} (\epsilon_{r\infty,33})$ & 3.12 (2.46) & 4.75 (2.30) & 5.41 (2.85) & 4.32 (3.46) & 4.93 (2.97) & 4.60 (2.23)  \\
\end{tabular}
\end{ruledtabular}
\end{table*}

\subsubsection{\label{sec:piezo}Piezoelectric properties}

All ferroelectric materials have piezoelectric properties, where an applied mechanical force results in a change in the spontaneous polarization, or conversely, an applied electric field results in a strain within the material. This coupling of mechanical and electric properties makes piezoelectrics applicable in a broad range of actuator and sensor devices. The piezoelectric charge coefficient $d_{ij}$ is often used to list the piezoelectric response of a material and is given by $d_{ij} = s_{jk}e_{ij}$, where $s_{jk}$ is the compliance tensor, and $e_{ij}$ is the piezoelectric stress coefficient. 

The piezoelectric coefficients $d_{ij}$ and the dielectric responses $\epsilon_{\mathrm{ii'}}$ of the identified OPTFe candidates are listed in Tab.~\ref{tab:table_piezo}. The longitudinal piezoelectric coefficients are modest, with magnitudes ranging from $0.05$ to $1.66~$pC/N. The magnitudes of the shear coefficients are somewhat larger and range from $0.49$ to $4.39$pC/N for all materials except YEKHUT. 
YEKHUT has a $d_{25} =-60.3~$pC/N, due to significantly larger shear compliance values than the other materials. Finally, we note that UHUMEP01 has a more anisotropic dielectric response than the other materials.

\section{\label{sec:conclusion} Summary and Conclusions }
By mining the CSD, we identified seven \mbox{OPTFes} candidates, in addition to 
20 earlier reported compounds.
Their spontaneous polarizations are found to range from 0.3 to 14.9 $\rm \mu$C/cm$^2$.
For earlier known compounds, we found that the experimental coercive fields
correlate with the PT barriers, and we used such linear fits to estimate the coercive field of the identified compounds, indicating values in the 1.2 to 18.0 kV/cm range.
The DFT calculations and the correlation analysis suggested that OPTFe tautomers have some degree of a collective PT switching nature. 
Our comparisons of compounds are in line with the intuitive notion that increasing the density of the PT-hydrogen bond can be used to increase the spontaneous polarization. 
While the identified compounds can themselves be of technological interest,
our study also provides inspiration for synthesizing new OPTFes.

\begin{acknowledgments}
The computations of this work were carried out on UNINETT Sigma2 high-performance computing resources (grant NN9650K). This work is supported by the Research Council of Norway as a part of the Young Research Talent project FOX (302362).
\end{acknowledgments}

\appendix

\bibliography{ref}
\end{document}